\documentclass[12pt,twoside]{article}
\usepackage[mathscr]{eucal}
\usepackage{amsmath,amsfonts,amssymb,amsthm,mathabx}
\usepackage{times}
\usepackage{pdfsync}
\usepackage{cite}
\usepackage{url}
\usepackage{hyperref}

\advance\textheight\topskip
\newcommand{\dd}{\partial}
\renewcommand{\d}{\partial}

\newcommand{\half}{\frac{1}{2}}

\newcommand{\ffrac}[2]{\raisebox{.5pt}%
  {\footnotesize$\displaystyle\frac{#1}{#2}$}\kern1pt}

\newcommand{\ddl}[2]{\ffrac{\dd #1}{\dd #2}}

\newcommand{\vddl}[2]{{\ffrac{\delta #1}{\delta #2}}}

\numberwithin{equation}{section} \makeatletter
\@addtoreset{equation}{section}

\DeclareFontFamily{OT1}{rsfs}{} \DeclareFontShape{OT1}{rsfs}{m}{n}{
<-7> rsfs5 <7-10> rsfs7 <10-> rsfs10}{}
\DeclareMathAlphabet{\mycal}{OT1}{rsfs}{m}{n}

\newcommand*\xbar[1]{%
  \hbox{%
    \vbox{%
      \hrule height 0.5pt 
      \kern0.3ex
      \hbox{%
        \kern-0.0em
        \ensuremath{#1}%
        \kern-0.0em
      }%
    }%
  }%
}

\begin{document}

\title{Conserved currents in the Cartan formulation
  of general relativity} 

\author{Glenn Barnich, Pujian Mao and Romain Ruzziconi}

\date{}

\def\mytitle{Conserved currents in the Cartan formulation of general
  relativity}

\pagestyle{myheadings} \markboth{\textsc{\small G.~Barnich, P.~Mao,
    R.~Ruzziconi}}{%
  \textsc{\small Conserved currents in the Cartan formulation of GR}}

\addtolength{\headsep}{4pt}

\begin{centering}

  \vspace{1cm}

  \textbf{\Large{\mytitle}}

  \vspace{1.5cm}

  {\large Glenn Barnich$^{a}$, Pujian Mao$^{b}$ and Romain Ruzziconi$^{a}$}

\vspace{.5cm}

\begin{minipage}{.9\textwidth}\small \it  \begin{center}
   ${}^a$ Physique Th\'eorique et Math\'ematique \\ Universit\'e Libre de
   Bruxelles and International Solvay Institutes \\ Campus
   Plaine C.P. 231, B-1050 Bruxelles, Belgium
 \end{center}
\end{minipage}

\vspace{.5cm}

\begin{minipage}{.9\textwidth}\small \it  \begin{center}
    ${}^b$ Institute of High Energy Physics\\
    and Theoretical Physics Center for Science Facilities \\
    Chinese Academy of Sciences\\ 19B Yuquan Road, Beijing 100049,
    P. R. China
 \end{center}
\end{minipage}

\end{centering}

\vspace{1cm}

\begin{center}
  \begin{minipage}{.9\textwidth}
    \textsc{Abstract}. We derive the expressions for the local,
    on-shell closed co-dimension 2 forms in the Cartan formulation of
    general relativity and explicitly show their equivalence to those
    of the metric formulation.
  \end{minipage}
\end{center}

\vfill

\noindent
\mbox{}
\raisebox{-3\baselineskip}{%
  \parbox{\textwidth}{\mbox{}\hrulefill\\[-4pt]}} {Proceedings of the
  workshop ``About various kinds of interactions'' in honour of
  Philippe Spindel,  4 \& 5 June 2015, Mons, Belgium}

\thispagestyle{empty}
\newpage

\section{Introduction}
\label{sec:introduction}

Surface charges in general relativity and gauge theories have a long
history that goes back to the founding papers on the Hamiltonian
formulation, see \cite{Arnowitt:1962aa} for a review and
\cite{Regge:1974zd} for further developments. Covariant approaches
based on the linearized theory are discussed in \cite{Misner:1970aa},
chapter 20, and also in \cite{Abbott:1981ff,Abbott:1982jh}. A
non-exhaustive list of subsequent references includes
\cite{Brown:1993br,Iyer:1994ys,Iyer:1995kg,Wald:1999wa,Anderson:1996sc}. More
recently, there has been interest in first order formulations, see
e.g.,
\cite{Julia:1998ys,Silva:1998ii,Julia:2000er,JacobsonMohd2015,%
CorichiReyes2015,CorichiRubalcava-GarciaVukasinac2016}.

Our approach here is based on actions, or more precisely, equivalence
classes of Lagrangians up to total divergences. It originates in
applications of the Batalin-Vilkovisky formalism to the perturbative
renormalization of gauge theories \cite{Barnich:1994db},
\cite{Barnich:2000zw}, but can also be formulated entirely
independently of this machinery, see
\cite{Barnich:2001jy,Barnich:2003xg,%
  Barnich:2004ts,Barnich:2007bf,Barnich:2010eb,Barnich:2011mi,Barnich:2013axa}
for details.

The aim of this note is to provide explicit expressions for the local,
on-shell closed co-dimension 2 forms in the Cartan formulation of
general relativity and prove their equivalence with those of the
metric formulation. The present note is extracted from a more complete
investigation that covers other first order formulations of general
relativity \cite{BarnichMaoRuzziconi2016}.

\section{Generalities}
\label{sec:generalities}

\subsection{Local BRST cohomology and generalized auxiliary fields}
\label{sec:local-bv-cohomology}

One of the virtues of the approach is that non-trivial, local,
co-dimension $2$ forms that are closed for all solutions of the
equations of motion can be shown to be isomorphic to local BRST
cohomology classes in ghost number $-2$. In turn, the latter are
naturally covariant under field redefinitions as well as suitably
invariant under the introduction and elimination of auxiliary and
generalized auxiliary fields \cite{Barnich:1994db}. Auxiliary fields
are a set of fields whose Euler-Lagrange equations of motion can be
solved algebraically to determine them in terms of the remaining
fields of the variational principle. Generalized auxiliary fields
extend this concept to the master action
\cite{DresseGregoireHenneaux1990,DresseFischGregoireEtAl1991}. They
are present whenever the vanishing of the gauge transformations of the
fields can be solved algebraically for some of the gauge
parameters. The associated generalized auxiliary fields are sub-sets
of fields which are algebraically pure gauge, in the sense that they
can be shifted arbitrarily by gauge transformations that do not
involve derivatives.

This is relevant for our purpose since the components of the Lorentz
connection in the Cartan formulation are auxiliary fields, while going
from the vielbein to the metric formulation involves elimination of
generalized auxiliary fields. Indeed, in the linearized formulation
the skew-symmetric part of the vielbein fluctuations are algebraically
pure gauge since they can be shifted arbitrarily by Lorentz
rotations. The argument can then be extended to the non-linear theory
as well, for instance by a perturbative analysis.

More details will be provided in \cite{BarnichMaoRuzziconi2016}.

\subsection{General case}
\label{sec:expl-constr}

Let $\phi^i$ denote the fields of the variational principle, $n$ the
spacetime dimension and ${\mathcal L}=L\, d^nx$ the Lagrangian times
the volume form. Here and below, we use the notation
\begin{equation}
(d^{n-p}x)_{\mu_1\dots\mu_p}=\frac{1}{p!(n-p)!}
\epsilon_{\mu_1\dots\mu_p\mu_{p+1}\dots
  \mu_n}dx^{\mu_{p+1}}\dots dx^{\mu_n},\label{eq:60}
\end{equation}
where the wedge product is omitted, $\epsilon_{\mu_1\dots\mu_n}$ is
completely antisymmetric and $\epsilon_{01\dots n-1}=1$. Let
$\delta_\epsilon \phi^i=R^i_\alpha(\epsilon^\alpha)$ denote a
generating set of non trivial gauge transformations. Under standard
regularity assumptions, one can then show that there is an isomorphism
between equivalence classes of local, on-shell closed co-dimension $2$
forms, with two such forms being equivalent if they differ on-shell by
an exact local form, and equivalence classes of reducibility
parameters $\bar f^\alpha[x,\phi]$ satisfying
$R^i_\alpha(\bar f^\alpha)\approx 0$, with two sets of reducibility
parameters being equivalent if they agree on-shell. In other words,
the classification of local, on-shell closed co-dimension 2 forms is
done through the classification of reducibility parameters, which is a
tractable problem.

The construction of the $n-2$ forms from the reducibility parameters
can be summarized as follows. For any $f^\alpha$, standard
integrations by parts allow one to write 
\begin{equation}
  R^i_\alpha(f^\alpha)\vddl{\mathcal{L}}{\phi^i}=f^\alpha
  R^{+i}_\alpha(\vddl{\mathcal{L}}{\phi^i})
  +d_H S_f, 
\label{eq:1} 
\end{equation}
for some weakly vanishing $n-1$ form 
\begin{equation}
S_f=S^{i\mu}_\alpha(\ddl{}{dx^\mu}\vddl{\mathcal{L}}{\phi^i},f^\alpha). \label{eq:4} 
\end{equation}
The $n-2$ form is then obtained by applying the contracting homotopy
$\rho_H$ for the horizontal differential of the variational bi-complex
\cite{Andersonbook,Olver:1993} 
\begin{equation}
  \label{eq:5}
  \{d_H,\rho_H\}\omega^p=\omega^p\ {\rm for}\ p<n. 
\end{equation}
to $S_f$,
\begin{equation}
  \label{eq:7}
  k_f=\rho_H S_f. 
\end{equation}
Indeed, the Noether identities associated to the generating set of
non-trivial gauge transformations are 
\begin{equation}
  \label{eq:6}
  R^{+i}_\alpha(\vddl{\mathcal{L}}{\phi^i})=0.
\end{equation}
For particular reducibility parameters that satisfy $R^i_\alpha(\bar
f^\alpha)=0$, \eqref{eq:1} reduces to $d_H S_{\bar f}=0$ so that
\eqref{eq:5} reduces to 
\begin{equation}
  \label{eq:8}
  d_H k_{\bar f}=S_{\bar f}\approx 0. 
\end{equation}
One can then proceed to show that $k_{\bar f}$ satisfies \eqref{eq:8}
also for general reducibility parameters (see \cite{Barnich:2001jy}
for details).

In this discussion, we have neglected non-trivial, identically
conserved currents, which are related to the topology of the bundle of
fields. We have thus neglected ``magnetic'' charges and concentrated
on the ``electric'' ones. The former can easily be incorporated when
taking into account the cohomology of the horizontal differential of
the variational bi-complex in lower form degrees, and more
specifically, in degree $n-2$ for the present case.

\subsection{Linearized theories}
\label{sec:linearized-theories}

For definiteness, let us take the example of the Einstein-Hilbert
action in metric formulation, where a generating set of gauge
transformations is given by the Lie derivative of the metric,
$\delta_\xi g_{\mu\nu}={\mathcal L}_\xi g_{\mu\nu}$. In spacetime dimension
$n\geq 3$, one can then show that $\xi^\rho[x,g]$ can be assumed not
to depend on the fields, so that reducibility parameters correspond to
Killing vectors. Since a generic metric does not admit Killing
vectors, there are no non-trivial conserved $n-2$ forms in general
relativity. In linearized gravity however, a generating set of gauge
transformations is given by
$ \delta_\xi h_{\mu\nu}={\mathcal L}_\xi\bar g_{\mu\nu}$, where
$\bar g_{\mu\nu}$ is the background solution around which one
linearizes the theory. There are then as many conserved $n-2$ forms as
there are Killing vectors of the background solution. Explicit
expressions are obtained by applying the construction described
previously, but now in the framework of the linearized theory. For
Einstein gravity, this has been done explicitly in
\cite{Barnich:2001jy}.

More generally, for gauge theories linearized around a solution
$\bar\phi^i$ with gauge transformations
$\delta_\epsilon\varphi^i=R^i_\alpha[x,\bar\phi](\epsilon^\alpha)$,
one can show \cite{Barnich:2003xg} that one may obtain the
$n-2$ forms of the linearized theory from the weakly
vanishing Noether current $S_f$ of the full theory through
\begin{equation}
  \label{eq:9}
  k_f[\delta\phi,\phi]=k^{\mu\nu}_f(d^{n-2}x)_{\mu\nu}=\frac{|\lambda|+1}{|\lambda|+2}
\partial_{(\lambda)}[\delta\phi^i\vddl{}{\phi^i_{((\lambda)\nu)}}\ddl{}{dx^\nu}
S_f], 
\end{equation}
by replacing $f$ by reducibility parameters of the linearized theory,
$\phi^i$ by the background solution $\bar\phi^i$ and $\delta\phi^i$ by
any solution $\bar\varphi^i$ of the theory linearized around
$\bar\phi^i$. Explicit expressions for the higher order Euler-Lagrange
derivatives can be found in \cite{Andersonbook} and \cite{Olver:1993};
our conventions and notations for multi-indices are summarized in the
appendix of \cite{Barnich:2001jy}. 

This construction is applicable in the case of Lagrangians that are of
finite, arbitrarily high order in derivatives. In case $S_f$ is of
second order in derivatives, which usually requires the Euler-Lagrange
equations of motion to be of second order as well, one needs the
higher order Euler-Lagrange operators up to order $2$,
\begin{equation}
  \label{eq:64}
  k_f[\delta\phi,\phi]=\frac{1}{2}\delta\phi^i\vddl{}{\phi^i_\nu}\ddl{}{dx^\nu}
  S_f+\frac{2}{3}\d_\sigma[\delta\phi^i\vddl{}{\phi^i_{\nu\sigma}}\ddl{}{dx^\nu}
    S_f]. 
\end{equation}
For theories for which $S_f$ is of first order in derivatives, only
the first higher order Euler-Lagrange operator is involved and reduces
to the partial derivative, so that the formula simplifies to
\begin{equation}
  \label{eq:10}
  k_f[\delta\phi,\phi]=\half \delta\phi^i\ddl{}{\phi^i_\nu}\ddl{}{dx^\nu}
S_f. 
\end{equation}
A first order formulation can always be achieved by introducing suitable
auxiliary and generalized auxiliary fields. 

For notational simplicity, we take units where the gravitational
constant is $G=(16\pi)^{-1}$. More standard choices correspond to
multiplying the action and forms below by $(16\pi G)^{-1}$.

\subsection{Asymptotics}
\label{sec:asymptotics}

The strategy to use the linearized theory at infinity with prescribed
asymptotics in order to define conservation laws in general relativity
is discussed in detail in \cite{Misner:1970aa}. 

Rather than trying to develop a theory for the asymptotic case, as
done for instance in \cite{Barnich:2001jy} for the ``asymptotically
linear'' case, one can take a more pragmatic point view that consists
in using the formula for the $n-2$ forms above, while substituting
asymptotic reducibility parameters and asymptotic solutions determined
by the fall-off conditions instead of exact ones determined by the
linearized theory. The approach is reminiscent of the one for current
algebras associated to broken global symmetries described in
\cite{jackiw:1985}. As a result, the currents are in general neither
integrable nor conserved. This is precisely what happens for general
relativity with asymptotically flat boundary conditions at null
infinity \cite{Wald:1999wa,Barnich:2011mi,Barnich:2013axa}.

\section{Application to the Cartan formulation of GR}
\label{sec:appl-cart-form}

\subsection{Cartan formulation}
\label{sec:cart-form-non}

Consider an $n$ dimensional spacetime with a 
moving, (pseudo-)orthonormal frame, 
\begin{equation}
e_a={e_a}^\mu\ddl{}{x^\mu},\quad e^a={e^a}_\mu dx^\mu, \label{eq:2}
\end{equation}
where ${e_a}^\mu{e^a}_\nu=\delta^\mu_\nu$,
${e_a}^\mu{e^b}_\mu=\delta_a^b$, and $\d_a f=e_a(f)$.
The structure functions are defined by
\begin{equation}
[e_a,e_b]={D^c}_{ab}e_c \iff de^a=-\half {D^a}_{bc}e^be^c.\label{eq:13}
\end{equation}
For further use, note that if ${\mathbf e}={\rm det}\,{e^a}_\mu$ then 
\begin{equation}
  \label{eq:82}
  \d_\mu(\mathbf{e}\,{e^\mu}_a)=\mathbf{e}\, {D^b}_{ba},
\end{equation}
We thus assume that there is a pseudo-Riemannian metric,
\begin{equation}
g_{\mu\nu}={e^a}_\mu \eta_{ab} {e^b}_\nu\label{eq:11},
\end{equation}
with a flat (Lorentz) metric in tangent space,
$\eta_{ab}={\rm diag}((-)1,1,\dots,1)$.  As usual, tangent space
indices $a,b,\dots$ and world indices $\mu,\nu,\dots$ are lowered and
raised with $g_{ab}$, $g_{\mu\nu}$, and their inverses, and converted
into each other using the vielbeins ${e_a}^\mu$ and their inverse.

Local (Lorentz) rotations are denoted by ${\Lambda_a}^b(x)$ with
${\Lambda_a}^b\eta_{bc}{\Lambda_d}^c=\eta_{ad}$, or equivalently,
${\Lambda^d}_b {\Lambda_a}^b=\delta_a^d$.  Under a combined frame
rotation and coordinate transformation, we have
\begin{equation}
{e'_a}^\mu(x')={\Lambda_a}^b(x){e_b}^\nu(x){\Lambda^\mu}_\nu(x), \label{eq:109}  
\end{equation} 
with ${\Lambda^\mu}_\nu=\ddl{{x'}^\mu}{x^\nu}$. 

In addition, assume that there is an affine connection defined by 
\begin{equation}
D_c e_a={\Gamma^b}_{ac}e_b\label{eq:12}, 
\end{equation}
and that metricity holds, 
\begin{equation}
D_a \eta_{bc}=0.\label{eq:22}
\end{equation}
This implies in particular that 
\begin{equation}
\Gamma_{abc}=-\Gamma_{bac}\label{eq:32},
\end{equation}
In terms of the Poincar\'e
algebra,
\begin{equation}
  \label{eq:16}
  [J_{ab},J_{cd}]=\eta_{bc}J_{ad}-\eta_{ac}J_{bd}-\eta_{bd}J_{ac}+\eta_{ad}J_{bc},
  \quad [J_{ab},P_c]=\eta_{bc} P_a-\eta_{ac}P_b, 
\end{equation}
one defines the Lorentz connection $\Gamma=\half \Gamma^{ab}J_{ab}$,
with ${\Gamma^{ab}}={\Gamma^{ab}}_{\mu} dx^\mu={\Gamma^{ab}}_{c}e^c$,
and $e=e^a P_a$. 

The torsion and curvature tensors are defined by 
\begin{equation}
{T}=T^a P_a=d{e} +[{\Gamma}, {e}],\quad R=\half R^{ab} J_{ab}=d{\Gamma}
+\half[{\Gamma},{\Gamma}]\label{eq:23},
\end{equation}
where the wedge product is omitted, and the bracket is the graded
commutator.

More explicitly,
$T^a=\half {T^a}_{bc} e^b e^c=de^a+{\Gamma^a}_b e^b$, so that 
\begin{equation}
{T^a}_{\mu\nu}=\d_\mu {e^a}_\nu-\d_\nu
{e^a}_\mu+{\Gamma^a}_{b\mu}
{e^b}_\nu-{\Gamma^a}_{b\nu}{e^b}_\mu,\label{eq:70}
\end{equation}
\begin{equation}
{T^c}_{ab}=2{\Gamma^c}_{[ba]}+{D^c}_{ba},\label{eq:14}
\end{equation}
where round (square) brackets denote (anti) symmetrization of enclosed
indices divided by the factorial of the number of indices involved.
In this case,
\begin{equation}
  \label{eq:36}
  \d_\mu({\mathbf e}\, v^\mu)={\mathbf e}\,
  (D_\mu+{e_b}^\nu\d_\mu{e^b}_\nu)v^\mu=D_\mu({\mathbf
    e}\, v^\mu),
\end{equation}
with $D_\mu v^\mu=\partial_\mu v^\mu$ for the Lorentz connection and
the definition
\begin{equation}
  \label{eq:73}
  D_\mu{\mathbf e}=\mathbf{e}\,({e_b}^\nu\d_\mu{e^b}_\nu).
\end{equation}
In particular, this implies that 
\begin{equation}
  \label{eq:94}
  D_\mu(\mathbf{e}\,{e^\mu}_a)=\mathbf{e}\,{T^b}_{ab}. 
\end{equation}
For the curvature components, ${{R}^a}_b=\half {R^{a}}_{bcd} e^c e^d=d
{\Gamma^a}_b+{\Gamma^a}_c{\Gamma^c}_b$, we have 
\begin{equation}
{R^f}_{c\mu\nu}=\d_\mu
  {\Gamma^f}_{c\nu}-\d_\nu {\Gamma^f}_{c\mu}
  +{\Gamma^f}_{d\mu}{\Gamma^d}_{c\nu}-{\Gamma^f}_{d\nu}{\Gamma^d}_{c\mu},\label{eq:71}
\end{equation}
\begin{equation}
{R^f}_{cab}=\d_a
  {\Gamma^f}_{cb}-\d_b {\Gamma^f}_{ca}
  +{\Gamma^f}_{da}{\Gamma^d}_{cb}-{\Gamma^f}_{db}{\Gamma^d}_{ca}-{D^d}_{ab}{\Gamma^f}_{cd}. 
\label{eq:15} 
\end{equation}
Furthermore, 
\begin{equation}
  \label{eq:20}
  [D_a,D_b]v_c=-{R^d}_{cab}v_d-{T^d}_{ab}D_dv_c. 
\end{equation}
Under a local frame rotation, we have 
\begin{equation}
  \label{eq:25}
  {e}'=\Lambda {e}\Lambda^{-1},\quad {\Gamma'}=\Lambda
  {\Gamma}\Lambda^{-1}+\Lambda d \Lambda^{-1},
\end{equation}
so that 
\begin{equation}
  \label{eq:26}
  {T'}=\Lambda{T}\Lambda^{-1},\quad {R'}=\Lambda{R}\Lambda^{-1}. 
\end{equation}
Defining $\Lambda=\mathbf 1+{\omega}+O(\omega^2)$, with
$\omega= \half {\omega^{ab}}J_{ab}$, 
$\omega^{ab}=-{\omega^{ba}}$, we have 
\begin{equation}
  \label{eq:35}
  \delta_\omega \Gamma=-(d\omega+[\Gamma,\omega])\iff \delta_\omega
  {\Gamma^{ab}}=-(d{\omega^{ab}}+{\Gamma^a}_c{\omega^{cb}}+{\Gamma^b}_c{\omega^{ac}}), 
\end{equation}
and also 
\begin{equation}
\delta_\omega e=[\omega,e]\iff \delta_\omega e^a={\omega^a}_be^b. \label{eq:49}
\end{equation}
Under a coordinate transformation, we have 
\begin{equation}
  \label{eq:58}
  {{e'}^a}_\mu={\Lambda_\mu}^\nu{e^a}_\nu,\quad
  {{\Gamma'}^a}_{b\mu}={\Lambda_\mu}^\nu{{\Gamma}^a}_{b\nu}, 
\end{equation}
and for ${x'}^\mu=x^\mu-\xi^\mu+O(\xi^2)$,
${\Lambda^\mu}_\nu=\delta^\mu_\nu-\d_\nu\xi^\mu+ O(\xi^2)$, so that
${\omega_\nu}^\mu=\d_\nu\xi^\mu$ and 
\begin{equation}
  \label{eq:59}
  \delta_\xi {e^a}_\mu={\mathcal L}_\xi e^{a}_\mu,\quad  \delta_\xi
  {\Gamma^a}_{b\mu}={\mathcal L}_\xi {\Gamma^a}_{b\mu}, 
\end{equation}
where ${\mathcal L}_\xi$ denotes the Lie derivative. 

The Bianchi identities are
\begin{equation}
  \label{eq:21}
  d{T}+[{{\Gamma}},{T}]=[{R},{e}],\quad d {R}+[\Gamma,R]=0. 
\end{equation}
Explicitly,
\begin{equation}
  \label{eq:24}
  {R^a}_{[bcd]}=D_{[b}{T^a}_{cd]}+{T^a}_{f[b}{T^f}_{cd]},\quad 
D_{[f}{R^a}_{|b|cd]}=-{R^a}_{bg[f}{T^g}_{cd]},
\end{equation}
where a bar encloses indices that are not involved in
the (anti) symmetrization. The Ricci tensor is defined by
${\mathbf R}_{ab}={R^{c}}_{acb}$, while
$S_{ab}={R^c}_{cab}=0$. Contracting the Bianchi identities gives
\begin{equation}
  \label{eq:27}
  {\mathbf R}_{ab}-{\mathbf R}_{ba}=-D_c {T^c}_{ab}
-2D_{[a} {T^c}_{b]c}-{T^c}_{fc}{T^f}_{ab},
\end{equation}
\begin{equation}
2D_{[f}{\mathbf R}_{|b|d]}+D_c{R^c}_{bdf}={\mathbf R}_{bg}{T^g}_{df}
-2{R^c}_{b[f|g|}{T^g}_{d]c}. \label{eq:28a}
\end{equation}
The curvature scalar is defined by
${\mathbf R}=g^{ab}{\mathbf R}_{ab}$, the Einstein tensor by
\begin{equation}
  \label{eq:33}
  G_{ab}={\mathbf R}_{ab}-\half g_{ab} {\mathbf R}. 
\end{equation}
Contracting \eqref{eq:28a} with $\eta^{bf}$ gives the contracted Bianchi
identities, 
\begin{equation}
  \label{eq:31}
  D_b {G^b}_a=\half {R^{bc}}_{da}{T^{d}}_{bc}+{{\mathbf
      R}^b}_c{T^c}_{ab}. 
\end{equation}

For any affine connection, metricity $D_{a}g_{bc}=0$, implies that
the connection is given by    
\begin{equation}
  \label{eq:17}
  \Gamma_{abc}=\{{}_{abc}\}+K_{abc}+r_{abc}, 
\end{equation}
where the Christoffel symbols are given by
\begin{equation}
\{{}_{abc}\}=\half(\d_b g_{ac}+\d_c g_{ab}-\d_ag_{bc})=\{{}_{acb}\}, \label{eq:97}
\end{equation}
$K_{abc}$ are the components of the contorsion tensor, 
\begin{equation}
  \label{eq:99}
K_{abc}=\half(T_{bac}+T_{cab}-T_{abc})=-K_{bac},
\end{equation} 
and 
\begin{equation}
  r_{abc}=\half(D_{bac}+D_{cab}-D_{abc})=-r_{bac}.\label{eq:96}
\end{equation}
Furthermore, one can directly show that 
\begin{equation}
  \label{eq:105}
  {\Gamma^a}_{b\mu}={e^a}_\nu(\d_\mu
  {e_b}^\nu+{\Gamma^\nu}_{\rho\mu}{e^\rho}_b)\iff 
{\Gamma}_{abc}=e_{a\nu}\d_c{e_b}^\nu+{e_a}^\mu{e_b}^\nu{e_c}^\rho\Gamma_{\mu\nu\rho}. 
\end{equation}
with 
\begin{equation}
  \label{eq:65}
  \Gamma_{\mu\nu\rho}=\{{}_{\mu\nu\rho}\}+K_{\mu\nu\rho}.
\end{equation}
Note also that for a Lorentz connection, \eqref{eq:17} reduces to 
\begin{equation}
  \label{eq:37}
  \Gamma_{abc}=K_{abc}+r_{abc}. 
\end{equation}

\subsection{Variational principle}
\label{sec:cartan-formulation}

In the standard Cartan formulation, the variables of the variational
principle are the components of the vielbein ${e_a}^\mu$ and a Lorentz
connection 1-form in the coordinate basis, ${\Gamma^a}_{b\mu}$ in
terms of which the action is
\begin{equation}
  \label{eq:69}
  S^C[{e_a}^\mu,{\Gamma^b}_{c\nu}]=\int d^nx\, L^C=\int d^nx\, {\mathbf e}\,
  ({R^{ab}}_{\mu\nu}{e_a}^\mu{e_b}^\nu-2\Lambda). 
\end{equation}
Using \begin{equation}
  \label{eq:47}
  \delta
  {R^a}_{b\mu\nu}=D_\mu\delta{\Gamma^a}_{b\nu}-D_\nu\delta{\Gamma^a}_{b\mu}, 
\end{equation}
the variation of the action is given by 
\begin{equation}
  \label{eq:74}
  \delta S^C=\int d^nx\, {\mathbf e}\,\big[2({G^a}_\mu
+\Lambda{e^a}_\mu)\delta {e_a}^\mu+{e_a}^\mu{e_b}^\nu
(D_\mu\delta {\Gamma^{ab}}_\nu-D_\nu\delta{\Gamma^{ab}_\mu})\big].  
\end{equation}
Using now \eqref{eq:36} and neglecting boundary terms, this gives
\begin{equation}
  \label{eq:75}
  \delta S^C=\int d^nx\, \big[2{\mathbf e}\,({G^a}_\mu
  +\Lambda{e^a}_\mu)\delta {e_a}^\mu+2 D_\nu({\mathbf e}\,
{e_a}^\mu{e_b}^\nu)\delta {\Gamma^{ab}}_\mu\big],
\end{equation}
so that 
\begin{equation}
  \label{eq:76}
  \vddl{L^C}{{e_a}^\mu}=2{\mathbf e}\,({G^a}_\mu
  +\Lambda{e^a}_\mu), 
\end{equation}
\begin{equation}
  \label{eq:77}
  \vddl{L^C}{{\Gamma^{ab}}_\mu}=2 D_\nu({\mathbf e}\,
{e_{[a}}^\mu{e_{b]}}^\nu)={\mathbf e}\, ({T^\mu}_{ab}+2e^\mu_{[a}{T^c}_{b]c}). 
\end{equation}
Contracting the equations of motions associated to \eqref{eq:77} with
${e_\mu}^b$ gives ${T^b}_{ab}=0$. When re-injecting, this implies
${T^a}_{bc}=0$. It follows that when the equations of motion for
${\Gamma^{ab}}_\mu$ hold, the connection is torsionless and thus given
by $\Gamma_{abc}=r_{abc}$. The fields ${\Gamma^{ab}}_\mu$ are thus
entirely determined by ${e_a}^\mu$ so that
${\Gamma^{ab}}_\mu$ are auxiliary fields.

Using \eqref{eq:74} for an infinitesimal gauge transformation as
in \eqref{eq:35}, \eqref{eq:49}, \eqref{eq:59} under the form 
\begin{equation}
  \label{eq:78}
  \delta_{\xi,\omega} S^C=\int d^nx\, \big[\vddl{L^C}{{e_a}^\mu}
\delta_{\xi,\omega} {e_a}^\mu
+\vddl{L^C}{{\Gamma^{ab}}_\mu}\delta_{\xi,\omega} {\Gamma^{ab}}_\mu\big],
\end{equation}
and integrating by parts in order to isolate undifferentiated gauge
parameters as in \eqref{eq:6} gives the Noether identities
\begin{equation}
  \label{eq:79}
  \vddl{L^C}{e^{[a|\mu|}}{e_{b]}}^\mu+D_\mu\vddl{L^C}{{\Gamma^{ab}}_\mu}=0,
\end{equation}
\begin{equation}
  \label{eq:80}
  \vddl{L^C}{{e_a}^\mu}\d_\rho {e_a}^\mu+\vddl{L^C}{{\Gamma^{ab}}_\mu}
\d_\rho{\Gamma^{ab}_\mu}+\d_\mu(\vddl{L^C}{{e_a}^\rho}{e_a}^\mu
-\vddl{L^C}{{\Gamma^{ab}}_\mu}{\Gamma^{ab}}_\rho)=0. 
\end{equation}
Equation \eqref{eq:79} can be shown to be equivalent to
\eqref{eq:27}. Using \eqref{eq:79}, equation \eqref{eq:80} 
can be written as
\begin{equation}
  \label{eq:83}
  \d_\mu(\vddl{L^C}{{e_a}^\rho}{e_a}^\mu)+\vddl{L^C}{{e_a}^\mu}D_\rho
  {e_a}^\mu
  +\vddl{L^C}{{\Gamma^{ab}}_\mu}{R^{ab}}_{\rho\mu}=0, 
\end{equation}
and then be shown to be equivalent to \eqref{eq:31}. 

\subsection{Construction of the co-dimension 2 forms}
\label{sec:constr-co-dimens}

When keeping the boundary term, one finds the weakly vanishing Noether
current associated to the gauge symmetries as 
\begin{equation}
  \label{eq:81}
  S^\mu_{\xi,\omega}=\vddl{L^C}{{\Gamma^{ab}}_\mu}(-{\omega^{ab}}
  +{\Gamma^{ab}}_\rho\xi^\rho)
  -\vddl{L^C}{{e_a}^\rho}{e_a}^\mu\xi^\rho. 
\end{equation}
The associated co-dimension 2 form
$k_{\xi,\omega}=k^{\mu\nu}_{\xi,\omega}(d^{n-2}x)_{\mu\nu}$ computed
through \eqref{eq:10} is
given by
\begin{multline}
  \label{eq:84}
  k^{\mu\nu}_{\xi,\omega}={\mathbf e}\,\big[(2\delta{e_a}^\mu
  {e_b}^\nu+{e^c}_\lambda\delta
  {e_c}^\lambda{e_a}^\nu{e_b}^\mu)(-\omega^{ab}+{\Gamma^{ab}}_\rho\xi^\rho)\\+
\delta
  {\Gamma^{ab}}_\rho (\xi^\rho {e_a}^\mu{e_b}^\nu 
+2\xi^{\mu} {e_a}^{\nu} {e_b}^\rho) -(\mu\longleftrightarrow\nu)\big]. 
\end{multline}
This can also be written as
\begin{equation}
  \label{eq:102}
  k_{\xi,\omega}=-\delta
  K^K_{\xi,\omega}+K^K_{\delta\xi,\delta\omega}-\xi^\nu\ddl{}{dx^\nu}\Theta_\xi, 
\end{equation}
where 
\begin{equation}
  \label{eq:87}
  K^K_{\xi,\omega}=2\mathbf{e}\,{e_a}^\nu {e_b}^\mu
  (-\omega^{ab}+{\Gamma^{ab}}_\rho\xi^\rho)(d^{n-2}x)_{\mu\nu},\quad 
\Theta_\xi= 2\mathbf{e}\,
\delta{\Gamma^{ab}}_\rho{e_a}^\mu{e_b}^\rho (d^{n-1}x)_\mu. 
\end{equation}

According to the general results reviewed in section
\ref{sec:generalities}, the co-dimension 2 form is closed, $d_H
k_{\xi,\omega}= 0$, or equivalently, 
$\d_\nu k^{\mu\nu}_{\xi,\omega}=0$, if ${e_a}^\mu,{\Gamma^{ab}}_\mu$
are solutions to the Euler-Lagrange equations of motion, and thus to
the Einstein equations, $\delta {e_a}^\mu,\delta {\Gamma^{ab}}_\mu$
solutions to the linearized equations and $\omega^{ab},\xi^\rho$
satisfy
\begin{equation}
  {\mathcal L}_\xi {e_a}^\mu+{\omega_a}^b{e_b}^\mu\approx 0,\quad
 {\mathcal L}_\xi {\Gamma^{ab}}_\mu\approx D_\mu 
  \omega^{ab}\label{eq:85}, 
\end{equation}
where $\approx$ now denotes on-shell for the background solution and
is relevant in case the parameters 
$\omega^{ab},\xi^\rho$ explicitly depend on the background solution
${e_a}^\mu,{\Gamma^{ab}}_\mu$ around which one linearizes. Note that
the first equation also implies in particular that $\xi^\rho$ is a
possibly field dependent Killing vector of the background solution
$g_{\mu\nu}$,
\begin{equation}
  \label{eq:88}
  {\mathcal L}_\xi g_{\mu\nu}\approx 0, 
\end{equation} 
and that 
\begin{equation}
  \label{eq:106}
\omega^{ab}\approx -{e^{b}}_\mu{\mathcal L}_\xi
{e^{a\mu}}\approx -{e^{[b}}_\mu{\mathcal L}_\xi
{e^{a]\mu}}. 
\end{equation}

\subsection{Reduction to the metric formulation}
\label{sec:reduct-metr-form}

In order to compare with the results in the metric formulation, let us
go on-shell for the auxiliary fields ${\Gamma^{ab}}_\mu$ and eliminate
$\omega^{ab}$ using \eqref{eq:106}. The former implies that we are in
the torsionless case with the Lorentz connection simplified to
${\Gamma^{ab}}_\mu={r^{ab}}_\mu$, while \eqref{eq:105} reduces to
\begin{equation}
  \label{eq:103}
  {\Gamma^{ab}}_\mu={e^a}_\nu\nabla_\mu
  e^{b\nu}={e^{[a}}_\nu\nabla_\mu
  e^{b]\nu}, 
\end{equation}
with $\nabla_\mu v^\nu=\d_\mu v^\nu+\{{{}^\nu}_{\rho\mu}\}v^{\rho}$.
Note also that the Killing equation can be written as
$\nabla_\mu\xi_\nu+\nabla_\nu\xi_\mu\approx 0$. 
Together with
\eqref{eq:103}, we have
\begin{equation}
-\omega^{ab}+{\Gamma^{ab}}_\rho\xi^\rho\approx
-{e^{[a}}_\rho{e^{b]}}_\sigma\nabla^\rho\xi^\sigma,\label{eq:90}
\end{equation}
\begin{equation}
  \label{eq:91}
  \delta{\Gamma^{ab}}_\rho=\delta {e^{[a}}_\sigma \nabla_\rho
    {e^{b]\sigma}}+ {e^{[a}}_\sigma \delta \{{{}^\sigma}_{\tau\rho}\}
      {e^{b]\tau}}+{e^{[a}}_\sigma \nabla_\rho \delta {e^{b]\sigma}}, 
\end{equation}
with
\begin{equation}
  \label{eq:92}
  \delta \{{{}^\sigma}_{\tau\rho}\}=\half g^{\sigma\delta}(
  \nabla_\rho\delta g_{\delta\tau}+\nabla_\tau\delta
  g_{\delta\rho}
  -\nabla_\delta\delta g_{\tau\rho}). 
\end{equation}
Using that 
\begin{equation}
  \label{eq:107}
  \delta {e^a}_\mu{e_a}_\nu=\half h_{\mu\nu}+\delta {e^{a}}_{[\mu}e_{|a|\nu]}, 
\end{equation}
with $h_{\mu\nu}=\delta g_{\mu\nu}$, indices being lowered and raised
with $g_{\mu\nu}$ and its inverse, and $h=h^\mu_\mu$, substitution
into \eqref{eq:84} gives
\begin{equation}
6\sqrt{|g|}\nabla_\rho(\delta {e_a}^{[\mu} e^{|a|\nu}\xi^{\rho]})+
  k^{\mu\nu}_\xi,\label{eq:3}
\end{equation}
where the first term can be dropped since it is trivial in the sense
that it corresponds to the exterior derivative of an $n-3$ form, while
\begin{multline}
  \label{eq:108}
  k^{\mu\nu}_\xi=\sqrt{|g|}\,\big[\xi^\nu\nabla^\mu h+\xi^\mu\nabla_\sigma
  h^{\sigma\nu}+\xi_\sigma \nabla^\nu h^{\sigma\mu}\\+\half
  h\nabla^\nu\xi^\mu+\half h^{\mu\sigma}\nabla_\sigma\xi^\nu+\half
  h^{\nu\sigma}\nabla^\mu\xi_\sigma-(\mu\longleftrightarrow \nu)\big].
\end{multline}
We have thus recovered the results of the metric formulation since the
last expression agrees with the one given in
\cite{Barnich:2003xg}\footnote{up to a typo in the second term of
  equation (35) in that reference, where
  $\tilde \xi^\mu D_\sigma h^{\sigma\mu}$ should read
  $\tilde \xi^\mu D_\sigma h^{\sigma\nu}$.}, which in turn is
equivalent to those derived directly in the metric formulation in
\cite{Barnich:2001jy}.

\section*{Acknowledgments}
\label{sec:acknowledgements}

\addcontentsline{toc}{section}{Acknowledgments}

This work is supported in part by the Fund for Scientific
Research-FNRS (Belgium) and by IISN-Belgium. The work of P. Mao is
supported in part by NSFC Grant No. 11575202.


\providecommand{\href}[2]{#2}\begingroup\raggedright\endgroup

\end{document}